\documentclass{aa}
\usepackage{graphicx}
\usepackage{txfonts}
\usepackage{natbib}
\usepackage{rotating}
\usepackage{color}
\usepackage{longtable,lscape}
\begin{document}
\title{A critical assessment of the metal content of the ICM}

\subtitle{}


\author{S. Molendi\inst{1},  D. Eckert\inst{2,1}, S. De Grandi\inst{3},
S.Ettori\inst{4}, F.Gastaldello\inst{1,5}, S.Ghizzardi\inst{1}, G.W.Pratt\inst{6} and M.Rossetti\inst{7,1} }

\offprints{S. Molendi \email{silvano@iasf-milano.inaf.it}}

\institute{INAF - IASF Milano, via E. Bassini 15 I-20133 Milano, Italy \and
Department of Astronomy, University of Geneva, Ch.d'Ecogia 16, 1290 Versoix, Switzerland \and
INAF - Osservatorio Astronomico di Brera, via E.Bianchi 46, 23807 Merate, Italy \and
INAF, Osservatorio Astronomico di Bologna, via Ranzani 1, 40127 Bologna, Italy \and
Department of Physics and Astronomy, University of California at Irvine, 4129 Frederick Reines Hall, Irvine, CA 92697-4575, USA \and
Laboratoire AIM, IRFU/Service d'Astrophysique - CEA/DSM - CNRS - Universit\'e Paris Diderot, B\^at. 709, CEA-Saclay, F-91191 Gif-sur-Yvette Cedex, France \and
Dipartimento di Fisica dell'Universit\`a degli Studi di Milano, via Celoria 16, I-20133, Milan, Italy
}
\date{\today}
\abstract
{}
{Our goal is to provide a robust estimate of the metal content of the ICM in
massive clusters.}
{We make use of published abundance profiles for a sample of $\sim$ 60 nearby systems, we include in
our estimate uncertainties associated to the measurement process and to the almost total
lack of measures in cluster outskirts.}
{We perform a first, albeit rough, census of metals finding
that the mean abundance of the ICM within r$_{180}$ is very poorly constrained,
0.06Z$_\odot$ $\lesssim$ Z $\lesssim$ 0.26Z$_\odot$, and presents no tension with expectations.
Similarly, the question of if and how the bulk of the metal content in clusters varies with cosmic time,
is very much an open one.}
{A solid estimate of abundances in cluster outskirts could be achieved by combining observations
of the two experiments which will operate on board Athena, the XIFU and the WFI, provided they do not
fall victim to the de-scoping process that has afflicted several space observatories over the last decade.}

\titlerunning{On the metal content of the ICM}

\authorrunning{Molendi et al.}

\maketitle

\section{Introduction}\label{sec:intro}
Over the last decade and a half, analysis of spectra from the latest generation of X-ray experiments,
has allowed the measure of   density,  temperature,  pressure
and entropy of the Intra-Cluster Medium (hereafter ICM) for several hundreds of systems.
For a more limited number of objects, long dedicated observations have permitted detailed studies of cores
and of other regions of particular interest.
More recently, the coming of age of SZ experiments has allowed  to construct cluster samples significantly
less biased and extending to greater cosmological distances than X-rays ones \cite[e.g.][]{planck29}.
All in all, the wealth of thermodynamic measures collected
out to redshifts of $\sim$ 1 has afforded an impressive improvement
in our understanding of the physics of these systems and made clusters one of the major
tools to estimate cosmological parameters.

X-ray data can also be used to derive another quantity, one that cannot be measured
with SZ experiments, and that has been used to a lesser extent than others, namely the metal abundance.
Spectra of high statistical
quality can be used to derive the abundance of several elements: i.e. O, Mg, Si, S, Ar,
Ca, Fe and Ni \cite[see][]{tamura04,mernier15}, however, for the  majority
of systems, measures are restricted to the most prominent line, i.e. the Fe K$\alpha$
line at 7 keV.
These measures have been used by several workers (see \citealt{deplaa13} for a recent review)
to infer several properties of the ICM. For example some have attempted to use metals
as tracers of gas motions in the ICM \cite[e.g.][]{ghizzardi13,rossetti10}.
A few  \cite[e.g.][]{tamura11} have audaciously attempted to use the limited spectral
resolution of X-ray CCD to directly measure shifts in the lines
\footnote{Things should be changing dramatically within a year from now with the launch of
the first X-ray micro-calorimeter, the Soft X-ray Spectrometer (SXS) on board the ASTRO-H
mission. With a spectral resolution of 6 eV, measurements of line shifts and broadenings
associated to subsonic motions are well within  the reach of the SXS.}.
However most of the effort thus far has gone into trying to characterise the metal
content of the ICM. Radial distributions of the metal abundance are available for several
tens of systems while 2D distributions have been published for a more limited number of
objects.
Several authors have compared abundances of different elements to point out that the
enrichment of the ICM requires
contributions from both SNIa and SNcc \citep{deplaa07,degrandi09}.
There have been a few attempts to connect the abundance distribution with the
cluster formation and evolution history \cite[e.g.][]{fabjan10}, indeed the amount of
metals that end up in the ICM is expected to depend critically upon the interplay between
star formation and AGN activity and can be used to provide constraints on feedback processes
that are complementary to the ones based on the entropy distribution \citep[e.g.][]{voit05,pratt10}.
 Some have attempted to gauge the enrichment process of the ICM \citep{tornatore07,cora08}
and to relate it to the nucleo-synthesis and ultimately the star formation
processes in cluster galaxies. More specifically, attempts to relate
the observed Fe content of the ICM with that expected from the stellar population,
have come to the conclusion that, for the most massive clusters, the
former exceeds the latter by a factor of several \cite[e.g.][]{loewenstein13}.
This is recognized as a problem because, unlike for less massive systems,
where the potential well is sufficiently shallow to allow the escape of at
least part of the metals injected into the ICM, for massive clusters all
metals are expected to remain within the system.
In this paper we provide a critical assessment of the metal content
of the ICM. In \S\ref{sec:abund} we briefly review the methodology and systematics involved
in measuring abundances in the ICM. In \S\ref{sec:metal} we perform our estimate of the
metal content of the ICM, while  in \S\ref{sec:disc} we discuss our finding and their implications.
In \S\ref{sec:fp} we consider future prospects for the measure of the Fe content of the ICM.
Finally in \S\ref{sec:sum} we summarize our results.

Abundances are reported relative to the solar photospheric values of \cite{anders89}, where
Fe$ = 4.68 \times 10^{-5} $ (by number relative to H). We make this choice despite the significant
evolution of solar reference systems of the last 2 decades,
(see \citealt{lodders10} for a review). Indeed, these changes have led to  the introduction of new references
which have been superseded in a matter of years. In this framework, the choice of an old but widely
adopted and recognized abundance reference is not such a bad one.
To ease comparison with other systems we recall that our Fe abundances can be converted
into the reference systems proposed in \cite{lodders03} and \cite{asplund09} by multiplying
respectively by 1.59 and 1.48.

Throughout this paper the term "metal"  will be used as a synonym for Fe.

\section{Measuring Abundances}\label{sec:abund}

Both continuum and line emission from the ICM are two body processes. The continuum emission is proportional
to the product of the number of all ions by the number of electrons, while the line emission is proportional
to the number of ions of a given species times the number of electrons. Under these conditions it is easy to
show that the equivalent width of a given line is proportional to the  ratio of the number density of the element producing
the line over the number density of hydrogen. In other words the equivalent width of the line is proportional to the relative abundance of the element producing it \cite[e.g.][]{sarazin88}.
Thus, even at the relatively modest resolution afforded by current CCD detectors,
$\sim 2\%$ at 6 keV, intense and isolated lines, such as the Fe${\rm K\alpha}$ line,
typically feature equivalent widths comparable or in excess of the spectral resolution.
 Several authors \cite[e.g.][]{deplaa13} have pointed out that the above  conditions
lead to direct and reliable estimates of the metal abundance in clusters.
Some year ago we \citep{degrandi09} carried out a  study to verify this expectation.
We performed a detailed spectral analysis of the spectra of the central regions  of 26 cool core clusters.
These systems are bright and the spectra are of high statistical quality allowing precise estimates of the equivalent
width of a few ion species.
We identified and investigated three possible causes of systematic uncertainties: 1) the calibration of the X-ray experiment;
2) the plasma code used to fit the X-ray data and 3) the thermal structure of the ICM.
By comparing measures of the three detectors onboard XMM-Newton we found that systematic errors on the abundance
to be below 3\% for Si and Fe. By comparing the {\tt mekal} \citep{mewe85,mewe86,kaastra92,liedahl95} and
{\tt apec} \citep{smith01_apec} plasma codes, available within the X-ray spectral
fitting package {\tt XSPEC} \citep{arnaud96_xspec}, we found differences of 10\% for Si and 5\% for Fe
\footnote{This may have improved over the last few years however, since
there is no single fitting package containing updated versions of these codes, the comparison is
somewhat complicated.}.
Finally, by fitting spectra with different multi-temperature models, namely 2T and 4T, we found that systematic differences
were always below 3\% for Fe.
So, in conclusion, focusing on Fe, which is the element that is typically used to measure the metal
abundance of the ICM, we found that systematic errors were below 5\%. Thus, we could confirm that, if data
of sufficient statistical quality is available, robust and precise estimates of the metal (i.e. Fe)
abundance can be made.

\section{Metal content of the ICM}\label{sec:metal}

The first measurement of the Fe line in clusters dates back to 40 years ago \citep{mitchell76}.
Since that time much has been learned about the metal content of clusters.
In the following subsection we review the relevant literature.

\begin{figure*}
\begin{center}
\includegraphics[width=7cm,angle=-90]{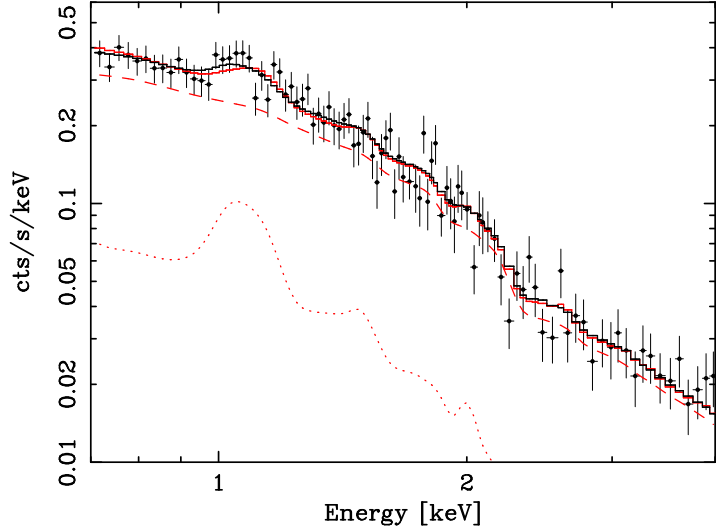}
\end{center}
\caption{\label{fig:spec} Simulated two temperature spectrum and best fitting one temperature model.
The simulated spectrum and its components are plotted in red, the 1.8 keV and 3.6 keV components are shown
as a dotted and dashed line respectively, while the total spectrum is reported as a solid line. The best
fitting one temperature model is shown as a solid black line. Both the simulation and the fit were performed
in a broad energy band, {\bf 0.5 keV - 10.0 keV}, however, here we show only the 0.7 keV - 4.0 keV range.
Note how the one temperature model
does an adequate job of reproducing the shape of the L-shell blend arising from the two temperature spectrum,
for a more detailed discussion see text and \cite{gasta10}.
Given the didactic nature of this simulation, background components have not been included, it should
go almost without saying that contributions from the latter will make attempts to discriminate between
different spectral models even more arduous.}
\end{figure*}

\subsection{The state of the art}\label{sec:mcont}
Amongst the first systematic
studies of abundance profiles is a paper based on BeppoSAX data, which we wrote more than a decade ago
\citep{degrandi01_ab}. There we showed for the first time that while non cool-core (hereafter NCC)
systems show relatively flat profiles, cool-core (hereafter CC) clusters feature an abundance excess in
their center. An important aspect of this measurement is that the profiles could be extended out to
about 0.4r$_{180}$
\footnote{r$_{180}$ is defined as the radius within which the mean density is 180 times the critical
density of the Universe. For massive clusters r$_{180}$ $\simeq$ 1.6r$_{500}$, where r$_{500}$ is another
reference radius often used in the literature, defined similarly to r$_{180}$.}.
More recently, we performed a more extensive study based on a sample of 60 massive clusters
\footnote{\bf In our works, we considered as "massive" systems with mean temperatures in excess of 3.5 keV
(see \citealt{leccardi08_t}) roughly corresponding to $3\times 10^{14}$M$_\odot$.}
observed with XMM-Newton \citep{leccardi10}. From Fig.6 of that paper we found that the strength of the metal abundance excess correlates with the central entropy in the sense that, within roughly 0.1r$_{180}$,
systems with lower central entropy feature a stronger central excess. In the intermediate region (i.e. between 0.1r$_{180}$ and 0.2r$_{180}$) the three groups of clusters we identified, namely low entropy core systems (LEC), medium entropy core systems (MEC) and high entropy core (HEC) systems all show an abundance
excess with respect to the metallicity measured in the outer region. Beyond 0.2r$_{180}$ we found no
evidence of any difference in abundance between clusters belonging to the three classes described above.

Another important point is
that this external region extends out to 0.4r$_{180}$. In other words, while the XMM-Newton observations have allowed to achieve a significant improvement in our characterization of core and circum-core regions, they have not permitted us to extend our exploration of the metal abundance of outskirts beyond what was previously known. We have discussed the reasons for this
failure elsewhere \citep{molendi04,ettori11} and we will not review them here. We do however wish
to draw our readers attention to the fact that measures of metal, i.e. Fe, abundance of clusters extend out to rather small radii. If we take as reference the mass of the ICM measured within r$_{180}$, as determined in \citet{eckert12}, see their Fig.6, we find that the gas mass within 0.1r$_{180}$, ranges between 2\% and 5\% of the total gas mass, depending on how peaked the surface brightness profile is in the core; it climbs to 10\% within 0.2r$_{180}$ and to 33\% within 0.4r$_{180}$. In other words the metal abundance of 2/3 of
the ICM is simply not known.

In recent years there has been one
\footnote{\bf There is also a measurement on the Virgo cluster \citep{simio15}, however this is a low mass
system with a total mass of roughly $1.4\times 10^14$ M$_\odot$ and we will not consider it further.} attempt to measure the metal abundance at larger radii with the Suzaku satellite on the Perseus cluster. \citet{werner13} measure a flat abundance profile out to $\sim$ 0.9r$_{180}$ with a mean value consistent with the one we determined in \citet{leccardi10}. This is certainly a very interesting measure, however it cannot be used to provide a robust estimate of the metal abundance in clusters at large radii for at least 4 good reasons.
1) The measure has been performed only on one system.
2) It does not provide a full azimuthal coverage of the cluster, indeed the coverage decreases as a function of radius.
3) The measure has been performed assuming a single temperature model, an assumption made in all the analysis of outer regions, including our own \citep{leccardi08_ab,leccardi10} but one that cannot be tested
with the current data and which could lead to systematic errors on the abundance {(see \S\ref{sub:bias} for details)}.
4) Even under the unverified assumption of a single temperature medium, abundance  measurements
in cluster outskirts are extremely challenging \cite[e.g.][]{ettori11} indeed, when the continuum from the cluster is only a few percent of the total signal that is measured, an exquisite characterization  of the background is required to provide reliable estimates of thermodynamic parameters. For the metal abundance, which typically requires the characterization of equivalent widths of a few hundreds eVs at 6 keV,
even more so.
This is illustrated by the fact that while a few tens of  measures of the temperature around r$_{180}$ are available in the literature \cite[e.g.][]{reiprich13}, there is only one for the abundance.

\begin{figure*}
\begin{center}
\includegraphics[width=7cm,angle=-90]{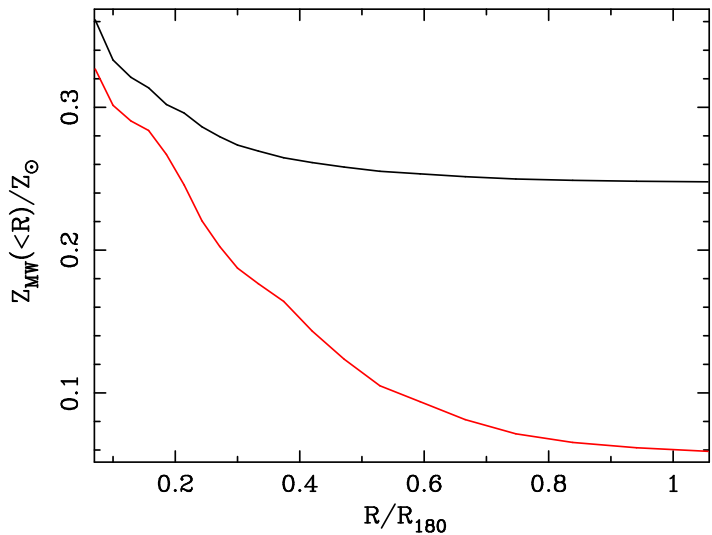}
\end{center}
\caption{\label{fig:abprof} Mean, cumulative, gas mass weighted, abundance profile for hot clusters.
The black and red lines trace, respectively, the rough upper and lower bounds of our estimate.}
\end{figure*}

\subsection{Abundance bias in the outskirts}\label{sub:bias}

In this subsection we provide an example to illustrate why metal abundances could be significantly
overestimated in cluster outskirts.
As pointed out in \S\ref{sec:mcont}, spectra from outer regions are typically fit with one temperature models, however there is mounting evidence that the temperature structure is more complicated.
Measurements of the surface brightness in outer regions indicate that inhomogeneities are certainly present on scales of several tens of kpc and beyond \citep{nagai11,roncarelli13,eckert15}, moreover, although
not yet detected, they likely extend to significantly lower sizes \cite[e.g.][]{gaspari14}.
Since regions of different density will be kept in approximate pressure equilibrium by sound waves,
the overdense regions will also be cooler and at a lower entropy than less dense regions.
A key issue is how metals are distributed between the different phases: do all
components share the same abundance or are some metal richer than others?
We have shown that in relaxed \cite[e.g.][]{ghizzardi13} and intermediate systems \citep{rossetti10}
entropy and metallicity tend to anti-correlate, i.e. regions of lower entropy are metal richer.
If  mixing and thermal equilibration processes are ineffective in homogenizing gas that is set into
motion by the continuous accretion process in clusters, as suggested by some of our recent work
\citep{eckert14,degrandi15}, it may well be that the lower entropy clumps
have different, likely higher, metallicity than the surrounding medium.

Hereafter we model in a rudimental fashion a multi-temperature plasma,
our purpose is not to reproduce in detail the thermodynamic structure of the ICM but more modestly
provide an example of how inhomogeneities characterized by the entropy
vs. metallicity anticorrelation could bias abundance measures.
We consider a volume of ICM which is filled with gas at two different densities, the more rarefied component filling most of the volume and the denser only 5\%.
Since, as already pointed out, sound waves will maintain gas of different densities in approximate pressure equilibrium, overdense region will also be cooler.
We assume the hotter component to be at 3.6 keV, a relatively low temperature similar
to that found in the outer regions of hot clusters and the cooler to be at half that temperature, i.e.
1.8 keV.  We also assume the cooler component to have a high metal abundance of 0.4Z$_\odot$  and the hotter one to have a lower metallicity of  0.05Z$_\odot$, in agreement with the scenario described above.
Since the cooler component, filling 5\% of the volume, is twice as dense as the hotter one, its mass will be 10\% of the total mass. We have simulated the spectrum from this two temperature plasma using the {\tt fakeit} command in the {\tt XSPEC}  spectral fitting package \citep{arnaud96_xspec} using  redistribution matrix and effective area files for the EPIC pn (we have also verified that, as expected, adopting response files from other CCD instrument does not change our results significantly).
By fitting the simulated spectrum with a one temperature model we derive a metal abundance in the range 0.2Z$_\odot$ - 0.25Z$_\odot$. This is a factor of $\sim$ 3 larger than the mass weighted metal abundance
that can be readily derived from the numbers
provided above, i.e. 0.08Z$_\odot$ {\bf and about a factor 2 larger than the emission weighted abundance, i.e. 0.11Z$_\odot$},  note that both simulated and fit spectra are shown in Fig.\ref{fig:spec}.
As pointed out in \S\ref{sec:mcont}, spectra from outer regions are typically fit with one temperature models mostly because the statistical quality of the data
is insufficient to allow multi temperature fitting. Moreover, even if a multi temperature
fit is attempted, some assumption on the relation between the metallicity of the different
components needs to be made as the metal abundance of the different components
are largely degenerate with respect to one another.  Fitting our simulated spectrum
with a two temperature model, with the two abundances free to vary independently of one another,
confirms this. If we assume that the two components have the same abundance we find a
metallicity of 0.13Z$_\odot$, almost a factor of 2 larger than the input mass weighted
abundance.

Despite some disagreement on the steepness of the temperature profiles,
 \cite[e.g.][]{walker12,eckert13a}, the shared view is that they
decline with radius. This implies that, with the  exception of the hottest
sytems, the Fe  abundance in the outermost regions of many clusters
will have to rely on L-shell measures alone. Under such circumstances, constraining
the metal abundance is equally if not more arduous.
In \citet{eckert14} we showed that the spectrum of a particular region could be fit
comparably well with a single temperature model and a multi-temperature one
differing by more than a factor of 3 in metal abundance.
Before closing this subsection, it is worth reminding our readers that significant biases in
the estimate of metal abundances in  multi-temperature regions have been recognized and
studied for some time \cite[e.g.][]{buote00b,buote00a,molendi01b,rasia08,gasta10}.
In his seminal papers Buote identified an "Fe-bias" which leads
to an underestimation of the Fe abundance, while in \cite{gasta10}
the authors describe an "inverse Fe-bias", working in the opposite direction, which
also happens to be the one at work in our example.
What is perhaps less well understood is that biases can be equally important in the outskirts as
in the cores of clusters \citep[see][for a discussion of this issue]{reiprich13}.


\begin{figure*}
\begin{center}
\includegraphics[width=12cm,angle=0]{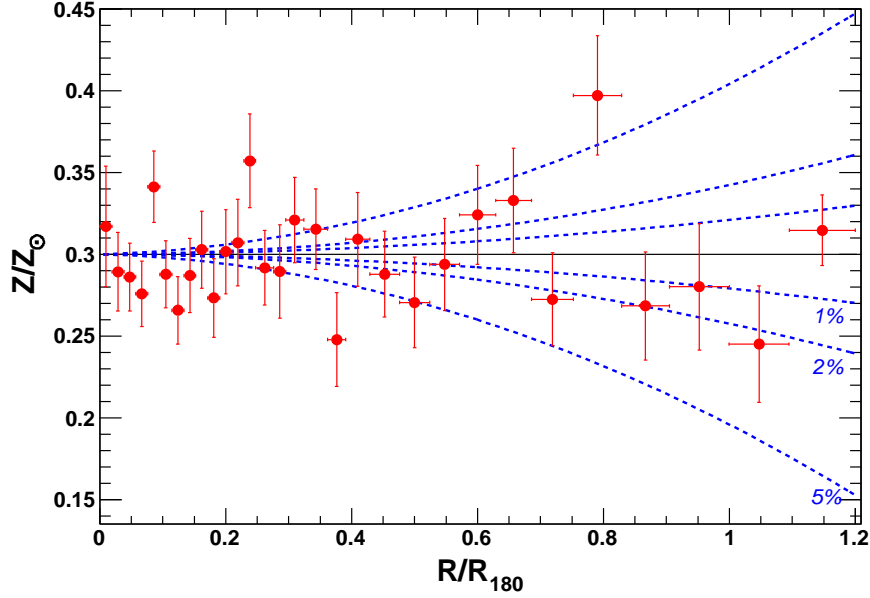}
\end{center}
\caption{\label{fig:sim} Fit of a simulated metal abundance profile for an 8 keV cluster
at $z=0.1$ for the WFI experiment on Athena. The simulated emission-measure
and temperature profiles are respectively from \citet{eckert12} and \citet{reiprich13}.
The spectra were simulated with a one temperature model and
the abundance was fixed to 0.3Z$_\odot$ everywhere. The effective area and background
intensity are those from the Athena proposal \citep{rau13}, the simulated exposure time is 100 ks.
The data points show the best fitting abundances and related statistical errors assuming perfect
reproducibility of the background,
the dashed lines refer to the systematic uncertainties associated to different levels of
reproducibility of the background, namely 1\%, 2\% and 5\%. Note how:  at small radii where the source
dominates over the background, systematic uncertainties on the latter are of little or no consequence;
at large radii, where the background dominates, its reproducibility drives the error budget.}
\end{figure*}

\subsection{A rough census}
In this subsection we attempt a first census of metals out to r$_{180}$.
As previously discussed, we have good estimates of the metal abundance within 0.2r$_{180}$ from
\cite{leccardi10}. For the region between 0.2r$_{180}$ and 0.4r$_{180}$ we also assume the abundance
 measured in \cite{leccardi10}, but we complement it with a factor of 2 uncertainty,
 see \S\ref{sub:bias}.
Since we do not have any reliable estimate of the metal abundance beyond 0.4r$_{180}$, we assume it to be anywhere
between 0.01Z$_\odot$, i.e. almost no metals\footnote{Although the measure on Perseus \citep{werner13}
could be significantly biased, the fact that the line has been detected shows that, however little, some Fe
must be present.}, and 0.24Z$_\odot$, the value measured between  0.2r$_{180}$ and 0.4r$_{180}$.
The upper bound is not based on any data but on the educated guess that metal abundance profiles do not increase with radius.
By combining these estimates with measures of the gas mass within a given radius \citep{eckert12} we derive
a mean, cumulative, gas mass weighted, metal abundance profile, Z$_{\rm MW}(<{\rm r})$, out to r$_{180}$,
see Fig\ref{fig:abprof}. The mean abundance within r$_{180}$, i.e. Z$_{\rm MW}(<{\rm r_{180}})$, is found to be anywhere between 0.06Z$_\odot$ and 0.26Z$_\odot$. We point out that, albeit weak, these are, to our knowledge, the first constraints on the metal content within r$_{180}$.

\section{Discussion}\label{sec:disc}

We have shown that the mean Fe abundance in clusters is very poorly constrained, ranging from a minimum of 0.06Z$_\odot$ to a maximum of 0.26Z$_\odot$. Although the data  used to performed the calculation
has been published for some time, and from our own group we might add, this is the first estimate
of the mean metal abundance within r$_{180}$. In this section we shall discuss, amongst other things, how,
despite their weakness, the limits on the metallicity can provide useful insight.

Over the last decades several authors have discussed the relation between the Fe content of the ICM, as estimated from observations, and the one expected from the stellar population observed in clusters \cite[e.g.][]{renzini97,loewenstein13,renzini14}.
Under the assumption that the most massive systems in our Universe, i.e. rich clusters, are closed boxes, in the sense that material that falls into them can no longer escape, their total Fe content should be easily predicted if SNe Fe yields are known and cluster  total stellar masses are measured \cite[e.g.][]{loewenstein13}.
The authors that have performed this exercise have found that the expected metal content of the ICM falls short of the observed one, assumed to be roughly 0.3 Z$_\odot$,  by a factor of several  \cite[e.g. 2 in][]{loewenstein13}.
In the previous section we have shown that, while the observed abundance is indeed  close to the one adopted by
these authors, the mean cluster abundance could differ from it very significantly, for the simple reason that the metallicity of the bulk of the ICM has yet to be measured.
For example in a recent paper \citep{renzini14}, derive a mean ICM metallicity of 0.3 Z$_\odot$, by making use of metal abundances measured  within 0.6 r$_{500}$, i.e. roughly 0.4 r$_{180}$ for rich clusters.

If taking an inventory of metals in local clusters is no easy task, trying to establish if
and how abundances vary across cosmic time is even more arduous. There have been several
attempts to characterize the redshift evolution of the metallicity \cite[e.g.][]{balestra07,maughan08,baldi12,ettori15}. In the latest and most complete of these papers,
\citet{ettori15}, we found no significant evidence of evolution in metallicity in the outermost regions,
roughly corresponding to the 0.2r$_{180}$-0.4 r$_{180}$ range. However, the question of if and how the bulk
of the metal content in clusters, which lies beyond our current reach, varies, remains very much an open one.

It is worth noting that there are other useful information that could be gleaned from the distribution of metals in outskirts. As already discussed in \S\ref{sec:intro}, these measures can be used to gauge
the formation and evolution process of cluster in a fashion that is complementary to the one
involving the measurement of  thermodynamic variables.  For example, it could be used to test AGN feedback processes (since metals efficiently trace bulk motions) and pinpoint the metal injection epoch \cite[e.g.][]{fabjan10}.

As an interesting aside, we note that in cluster cores  the metal budget does not appear to be a problem.
Under the assumption that the metal excess  observed in cool cores is due to stars currently residing
in the BCG invariably found at the center of these systems \citep{degrandi04}, we have found that the
measured Fe mass in the ICM is comparable to the expected one \cite[e.g.][]{degrandi14}.


\section{Future Prospects}\label{sec:fp}
The characterization of metal abundances in cluster outskirts requires data from new experiments that are
more sensitive to low surface brightness emission than existing ones.
The Japanese satellite ASTRO-H \citep{taka12} should be operational within the next year; it carries several experiments, the Soft X-ray Imager (SXI) is  comparable to the
CCDs on board Suzaku, with one notable exception, i.e. the significantly larger field of view,
$35^\prime \times 35^\prime$, and may lead to some improvements with respect to the latter experiment. The Soft X-ray Spectrometer (SXS) provides unprecedented spectral resolution and will undoubtedly make very significant contributions in several fields \citep[e.g.][]{kita14}. In the case at hand however, the  relatively high instrumental background combined with the small field of view, $3^\prime \times 3^\prime$, and effective area, 300 cm$^2$ at 6 keV, will make exploration of cluster outer regions with the SXS  challenging. Indeed no systematic study of cluster outskirts with SXS is
foreseen at this time \citep{kita14}.
The launch of Spektr-RG  is  currently scheduled for 2017; the eROSITA experiment \citep{merloni12},
on board Spektr-RG, has characteristics similar to previous CCD imagers and should have a sensitivity
to low surface brightness emission comparable to that of  the XMM-Newton EPIC cameras.

The ESA mission Athena, currently in its phase A study and with an expected  launch in 2028, will carry a
Wide Field X-ray Imager (WFI) and an X-ray Integral Field Unit (XIFU).
The XIFU is a microcalorimeter array with large collecting area and a $3.5^\prime \times 3.5^\prime$
field of view. The design of XIFU on board Athena includes an anti-coincidence
system as well as a passive shield, which are expected to reduce the instrumental background by
more than an order of magnitude \citep{lotti14}.
In the current design the WFI enjoys a combination of very large effective area, 2 m$^2$ at 1 keV, and
good angular resolution, $\sim 5^{\prime\prime}$, extending over a very large, $40^\prime \times 40^\prime$, field of view. Moreover, there are a  number of  current activities on the telescope and detector design that, if successful, will lead to an experiment with low, stable and extremely well characterized background.

While the limited field of view of the XIFU will likely not allow a full coverage
of cluster outer regions, its high spectral resolution combined with the low background will
afford a sampling of the temperature structure of cluster outskirts. This information can
be subsequently fed into the WFI spectral modelling: as pointed out in \S\ref{sub:bias},
an adequate characterization of the temperature structure of the plasma is a key point to keep biases in the abundance measurements under control.
In Fig.\ref{fig:sim}, we provide an example of an abundance profile measurements with the WFI
on board Athena, details on the properties of the simulated cluster are provided in the figure caption.
As shown by the figure, the profile can be measured out to about r$_{180}$. A key point is that
the fitting of the WFI spectra has been performed with the same spectral model that has
been used to simulate the data, in the case at hand a one temperature plasma.
Another point worth bearing in mind is that these measurements depend
upon a few critical parameters: a decrease in effective area or, alternatively, an increase
in background intensity, will lead to larger errors, that, assuming no change in the level of systematics, cannot be recovered with longer exposure times.
Similarly a loss of reproducibility of the background  will  lead to an increase of systematic errors,
this is illustrated in Fig.\ref{fig:sim}, where we show how rapidly systematic uncertainties grow
in the outskirts as background reproducibility degrades. For instance at r$_{500}$, corresponding to
$\sim$ 0.62r$_{180}$, we expect a systematic uncertainty on the measure of the abundance of
2\%, 4\% and 10\%  for a background reproducibility good to 1\%, 2\% and 5\% respectively,
while at r$_{180}$ we expect a systematic uncertainty of 7\%, 14\% and 35\% for the same levels
of background reproducibility.
Thus, a de-scoping of the mission entailing a loss in effective area, or in
background reproducibility, or an increase in background intensity, could limit significantly
measures of metal abundances in cluster outskirts.



\section{Summary}\label{sec:sum}

Our results are summarized as follows.

\begin{itemize}

\item Robust estimates, characterized by systematic uncertainties of the order of a few percent,
are available for cluster core and circum-cores. The mean metal
abundance within these regions, i.e. r $<$ 0.2r$_{180}$, is $\sim 0.3$ Z$_\odot$.

\item  Between 0.2r$_{180}$ and 0.4r$_{180}$ we have several tens of  measurements,
however the limited statistical quality of the data does not allow a detailed reconstruction
of the thermodynamic structure of the gas in these regions. This leads to systematic uncertainties
in metal abundance measures which could be as large as a factor of 2 or even more.

\item Since almost nothing is known of the metal abundance beyond 0.4r$_{180}$,
where the bulk of the gas mass resides, any claim of tension between the measured and expected metal
content of the ICM should be viewed with a good dose of skepticism.
Using published data and assuming that between 0.4r$_{180}$ and r$_{180}$ the abundance be constrained
between 0.016Z$_\odot$ and 0.246Z$_\odot$, we have performed a first, albeit rough, census of metals
within r$_{180}$.
We find the mean, mass weighted, abundance within r$_{180}$, i.e. Z$_{\rm MW}(<{\rm r_{180}})$,
to be between 0.06Z$_\odot$ and 0.26Z$_\odot$; this broad range does not conflict
with expectations.
Similarly, the question of if and how the bulk of the metal content in clusters varies with cosmic time,
is very much an open one.

\item A solid estimate of abundances in cluster outskirts could be achieved by combining observations
of the two experiments which will operate on board Athena, the XIFU and the WFI, provided they do not
fall victim to the de-scoping process that has afflicted several space observatories over the last decade.



\end{itemize}

\acknowledgements
S.M. would like to acknowledge the stimulating environment of the workshop:
``The Metal Enrichment of Diffuse Gas in the Universe'' held in Sesto Pusteria in
late July 2015, where the original idea for this paper was conceived.
We acknowledge financial contribution from contract PRIN INAF 2012 (”A unique dataset
to address the most compelling open questions about X-ray galaxy clusters”).


\bibliographystyle{aa}
\bibliography{biblio_Aug2015}

\end{document}